\documentstyle{article}

\begin{document}
%\draft
%\title
\centerline{ CAUSALITY, SYMMETRIES AND QUANTUM MECHANICS}
%\author
\centerline{Jeeva Anandan}
\bigskip
%\address
\centerline{ Clarendon Laboratory, University of Oxford}
\centerline{Parks Road, Oxford OX1 3PU, UK}
\smallskip
\centerline{and}
\smallskip
\centerline{ Department of Physics and Astronomy}
\centerline{University of South Carolina} \centerline{Columbia, SC
29208, USA} \centerline{E-mail: jeeva@sc.edu}
\bigskip
\centerline{December 8 , 2001, revised July 31, 2002}
%\maketitle
\bigskip
\bigskip
\begin{abstract}
It is argued that there is no evidence for causality as a
metaphysical relation in quantum phenomena. The assumptions that
there are no causal laws, but only probabilities for physical
processes constrained by symmetries, leads naturally to quantum
mechanics. In particular, an argument is made for why there are
probability amplitudes that are complex numbers, which obey the 
Born rule for quantum probabilities. This argument
generalizes the Feynman path integral formulation of quantum
mechanics to include all possible terms in the action that are
allowed by the symmetries, but only the lowest order terms are
observable at the presently accessible energy scales, which is
consistent with observation. The notion of relational reality is
introduced in order to give physical meaning to probabilities.
This appears to give rise to a new interpretation of quantum
mechanics.
\end{abstract}

\bigskip
\bigskip
\noindent
arXiv: physics/0112020.

\noindent
Published in 
{\it Foundations of Physics Letters,} Vol. 15, No. 5, 415-438 (October 2002).

\newpage
\section{Introduction}

The world-view that the universe is governed by precise causal or
dynamical laws, which was called the paradigm of laws
\cite{an1999}, was due to Galileo, Descartes, Newton and others.
It was argued, however, that the fundamental laws of physics are
effective laws arising from symmetries \cite{an1999}. This view
has the advantage of naturally giving rise to the indeterminism of
quantum phenomena. In the present paper, specific arguments
towards obtaining quantum theory in the absence of causal laws
using symmetries will be presented.

In section 2, the notion of causal laws will be critically
examined in the context of wave-particle duality of quantum
phenomena, from a historical point of view. A heuristic principle
that eliminates metaphysical relations in physics that are not
invariant under symmetries is formulated in section 3. This
principle implies that causality as a metaphysical relation should
be eliminated in a theory that has time-reversal symmetry. Then,
in section 4, the EPR correlation, indeterminism of the outcomes
of measurement, and the accepted view that there is no causal
relation between space-like separated events are used to argue
that there is also no metaphysical causal relation between
time-like separated events. However, physical causality that is
the ability to communicate information probabilistically
distinguishes these two pairs of events. This distinction is made
in section 5, and an argument, due to Weinberg, that the
non-causality between space-like separated events is due to the
Lorentz boost symmetries is presented. This reinforces the present
view on the fundamental role of symmetries. 

In sections 6 and 7, the physical causality is explained on the
basis of two postulates, namely that there are no causal laws but
that the probabilities of physical processes are invariant under a
group of symmetries. The first postulate naturally leads, on using
an algebraic theorem due to Horwitz and Albert, to probability
amplitudes as complex numbers. The Born rule is then derived, in
section 7, by considering a `which way' experiment in which
interference between probability amplitudes is lost. The second
postulate then yields a generalization of Feynman path integral
formulation in which the action includes all terms that are
invariant under the symmetries; but only the lowest order terms
that are in the standard model are observed in the experiments
performed with presently accessible energy scales. The question of
when a probability amplitude should be converted to probability,
which is called a quantum `measurement', is considered in section
8. The notion of relational reality that is jointly realized when
two systems interact is introduced. This enables the predictions
of probabilistic outcomes during `measurements,' while preserving
the universal validity of quantum mechanics. Also, this makes
quantum mechanics both local and realistic; but the realism is
relational instead of absolute, which enables the present view to
be consistent with Bell's theorem. The present interpretation is
shown to be intermediate between the Copenhagen and Everett
interpretations, and some speculations are made, in the final
section.

\section{Causal Laws and Wave-Particle Duality}

The first indications
that the paradigm of laws may not be valid already appeared during
Newton's study of light. Newton had
formulated a highly successful set of laws for material
particles, known today as Newton's laws of motion and
gravitation. Contrary to Newton's famous statement,
``hypotheses non fingo'', the assumption that the material
particles should obey these laws, or any laws, was a
hypothesis. It was natural for Newton to then try to bring the
behavior of light into this paradigm, which he has helped to
create more than anyone else up to that time. So, he made the hypothesis that
light consisted of material particles, called corpuscles.
This was convenient for Newton because then these corpuscles
are subject to the same laws of motion which he has
already `perfected'.

However, Newton's theory of light ran into problems almost
immediately. It could not explain partial reflection, as
Newton himself recognized. Why is that when a corpuscle
encountered a slab of glass, it was sometimes transmitted and
sometimes reflected? To explain this using the deterministic
laws of motion of Newton, it was supposed at first that there
are `holes' and `spots' in the glass, so that if the
corpuscle encountered a hole it went through and if it struck
a spot it was reflected. This was perhaps the first of the
patchwork theories of physics that were proposed in order for
physicists to remain within the paradigm of laws, which will
be done all the way to the 21st century. But Newton himself
realized that this theory does not work. This was because, as
Feynman \cite{fe1985} has mentioned, Newton made his own
lenses and mirrors by polishing glass. And he knew that the
small scratches that he made with powder as he polished glass
had no appreciable effect on the partial reflection of light.

Newton's hypotheses, in addition to not explaining partial
reflection, could not explain also interference and
diffraction, as is well known. Physicists tried to solve
these problems by abandoning Newton's ontology of corpuscles,
while keeping his basic assumption that light obeyed
deterministic laws. They should have done the reverse! But I
shall follow the historical route before discussing the
logical alternative. Historically, physicists {\it replaced}
Newton's corpuscles with a wave. What made this appealing to them was
that Huygens had formulated a law for the propagation of a
wave, called Huygens' principle, according to which every
point on a wave front acted as a source of secondary wavelets
whose interference was sufficient to reconstruct the
subsequent wavefronts. This was the first dynamical or causal law to
govern the propagation of a wave, as opposed to Newton's laws
that governed the propagation of material particles. And
using these laws and the new ontology that light {\it is} a
wave it was easy to explain all phenomena of light known at
that time, including partial reflection, interference and
diffraction.

But today we know additional phenomena that would make
Newton's ontology of light consisting of corpuscles appear to
be fundamentally valid. For example, if we make the intensity
of light falling on a photographic plate low enough we see
spots appearing, which is interpreted as due to the
corpuscles of light, or photons as they are now called,
interacting with the plate.
Newton could have saved his ontology for light by giving up
his hypothesis that the corpuscles should obey causal
deterministic laws. Suppose in the above example of the glass
slab, $30\%$ of the light is reflected and $70\%$ is
transmitted. Newton could have postulated that a corpuscle
has a $30\%$ {\it probability} of being reflected and
detected in a detector and $70\%$ probability of it being
transmitted and detected in a different detector. But
physicists were unwilling to give up causal
deterministic laws until the twentieth century when observed
physical phenomena made them question their cherished
beliefs.

The wave theory received a tremendous boost in the nineteenth
century with the introduction of electric and magnetic fields
by Faraday and Maxwell. These fields obeyed causal
deterministic laws that were mathematically formulated by
Maxwell. Moreover, light waves were recognized as special
cases of this electromagnetic field, and Maxwell's laws
justified Huygens' principle. The price paid now for staying
within the paradigm of laws was only that the universe had to
be regarded as a strange mixture of material particles and
fields. Physicists lived with this dual ontology even when an
inconsistency was found between the two sets of laws that
governed material particles and fields. This inconsistency,
first clearly recognized by Einstein, was that the symmetries
of the laws of mechanics that governed material particles
were not the same as the symmetries of the laws of the
electromagnetic field. Einstein required that both symmetries
should be the same, and asserted the primacy of fields over
particles by requiring that the laws of mechanics should be
modified so that they have the {\it same} Lorentz group of
symmetries as the laws of
the electromagnetic field. This was the first time in the
history of physics that symmetries took priority over laws in
the sense that the laws were modified to conform to the
symmetries. Moreover, the existence of universal symmetries
for all the laws of physics enabled the construction of a
physical geometry having the same symmetries, namely the
Minkowski space-time.

\section{Role of Symmetries in Eliminating Metaphysical Relations}
\label{metaphysical}

The Lorentz group of symmetries also eliminated the following three
metaphysical relations that existed prior to Einstein's paper
on relativity. 1) Newton's postulated
``absolute space'' (on the basis of his rotating bucket
argument), or the ``ether'' in which light waves propagated,
implied an absolute relation between two time-
like separated events that have the same {\it absolute position} in
this ``absolute space''. 2) Newtonian physics assumed that two events have the
relation of {\it absolute
simultaneity} if they occur at the same ``absolute time''.
3) Newtonian physics allowed for
causal relations to exist between absolutely simultaneous
events. For Newton's gravitational interaction,
these were the only causal relations. But since
these three relations are
not Lorentz invariant (relation (1) is not even
Galilei invariant), and therefore not an objective property
of the world, they were discarded. The overthrow of (3),
which Newton himself regarded as unnatural, meant that,
since any pair of space-like events is
simultaneous in an appropriately chosen inertial frame, and
all inertial frames are related by the Lorentz group of
symmetries, the resulting acausality needed to be extended to all
space-like separated events.

In order to eliminate such metaphysical relations in general,
I now formulate a heuristic principle, called $M$: {\it A necessary
and sufficient condition for a relation to be admissible as an ontological
relation in a physical theory is that it should be invariant
under all the symmetries of the theory}. Here the term `relation' is understood
in the usual sense of this term between two physical concepts or objects, in
general. The above three relations, specifically, are between events in space-
time $\bf S$. Any relation in $\bf S$ is defined mathematically as a subset of
${\bf S}\times {\bf S}$.
On the basis of $M$, the metaphysical relation (1) of absolute
position, and hence absolute space, was not admissible even
in Newtonian physics because this relation is not invariant
under the Galilei boosts that are symmetries of this theory.
But the other two metaphysical relations mentioned above were
invariant under the Galilei group of symmetries and therefore
excluded only because the Galilei group was superseded by
the Lorentz group of symmetries as a result of the work of
Lorentz, Poincare and Einstein.

However, there still remained the following metaphysical relation:
4) The causal relation between two events that are time-
like separated. But this relation is asymmetric in time
because while an earlier event $a$ may influence a later
event $b$, it is not possible for the later event to
influence the earlier event. Even in a deterministic theory
like classical electrodynamics, time asymmetric causality is
introduced by the choice of retarded Green's functions. This
is unlike Newton's causal relation between simultaneous
events, which is a symmetric relation because of Newton's
third law of motion, and invariant under the
Galilei group of symmetries. The physical theories we have
today are invariant under time reversal symmetry T, apart from
weak interaction which is irrelevant to the problem at hand because causality is
posited today even in the absence of weak interaction.
Moreover, all theories of physics today, including weak
interactions, have CPT symmetry. Hence, the principle
$M$, and time reversal symmetry in the absence of weak
interactions or CPT symmetry in the presence of weak
interactions imply that the metaphysical relation (4) of
causality should be discarded. Alternatively, if causality is to be kept then
CPT symmetry or T symmetry in the absence of weak interactions should be
discarded. But while people would agree that the later event $b$ cannot
influence the earlier event $a$, they would not accept the reverse even though
the intervening space-time region between $a$ and $b$ has the
same structure for both relations!

\section{Indeterminism and the EPR Paradox}
\label{indeterminism}
%reference to epr

Eliminating causality means that a given event is not
uniquely determined by the `earlier' events. Then there must be indeterminism in
our physical theory.
But historically, physicists seriously entertained indeterminism only after the
wave-particle duality was forced upon them.
The work of Planck and Einstein showed that it was necessary
to associate particles with the electromagnetic field,
vindicating the ontology of Newton. But at the same time, the field included
the wave aspect. This wave-particle duality was
recognized as characteristic of all particles by De Broglie,
and Schr\"odinger introduced the wave function that obeyed
Schr\"odinger's equation to represent the wave properties.
The relation between the wave and the
particle was given in probabilistic terms by Born: The
probability density of observing a particle at $\bf x$ is
$|\psi({\bf x})|^2$, where $\psi$ is the wave function of the
particle representing its state. More generally, the
probability of observing this particle in a state $\phi$ is
$|<\psi|\phi>|^2$. Originally, this Born rule was associated with an assumed
indeterministic change from the state $\psi$ to $\phi$.
Subsequently, two different deterministic descriptions of
quantum phenomena were given by Bohm \cite{bo1952} (also called the causal
interpretation of quantum mechanics) and
Everett \cite{ev1957}. But for these descriptions to be
empirically relevant, they need to give the experimentally
well confirmed Born rule. In the latter two
descriptions, however, probabilities are introduced ad hoc,
which amount to bringing the indeterminism of quantum
mechanics through the back door.

I shall now show that, owing to this indeterminism, the
correlation between two entangled states, between which there is no causal
relation, is metaphysically similar to the correlation
between two states that are related by Schr\"odinger evolution. Consider two
spin-half particles $1$ and $2$, which interacted some time ago but
no longer interacting, and whose spin states are now EPR
correlated:
\begin{equation}
\psi = {1\over \sqrt{2}} ( \psi_\uparrow(1)
\psi_\downarrow(2) - \psi_\downarrow(1) \psi_\uparrow(2))
\label{epr}
\end{equation}
where $\psi_\uparrow(1)$ is the spin-up state of particle $1$
and $\psi_\downarrow(2)$ is the spin-down state of particle
$2$ etc. The state (\ref{epr}) is spherically symmetric
because its total spin is zero; therefore the `spin-up' and
`spin-down' basis states may be with respect to any direction
in space. As is also well known, it is not possible to
communicate using the entanglement between the two particles
in (\ref{epr}) because they are non interacting.
Suppose Alice and Bob make measurements on
particles $1$ and $2$, respectively, and try to use their
outcomes to communicate. If Alice observes $1$ to have spin-
up (spin-down) along the $z-$direction, she can predict with
certainty that $2$ has spin-down (spin-up). Therefore, Bob by
measuring the spin along the $z-$direction for $2$ can
verify the outcome of Alice's measurement, provided Alice has
informed Bob beforehand that she will measure spin in the
$z-$direction. But it is impossible for Alice to send a
signal this way because of the {\it indeterminacy} of the
outcome of her own measurement. Suppose now that Alice and
Bob have decided beforehand that if Alice measures spin
along the
$z-$direction ($x-$direction) the signal she sends to Bob is
`yes' (`no'). But it is impossible for Bob to know which
observable Alice has actually measured because of the {\it
indeterminacy} of the outcome of his own measurement. Thus
the indeterminacy of quantum mechanics prevents Alice
communicating with Bob using the entanglement in (\ref{epr}).

The inability to communicate signals faster than the speed
of light $c$, and the ability to communicate signals with
speed less or equal to $c$ is called Einstein causality. We
saw above that since quantum mechanics allows for
entanglement, the indeterminism in the outcome of
measurements is essential to preserve Einstein causality.
People often wonder why non relativistic
quantum mechanics should preserve Einstein causality, which
was obtained from relativistic physics. However, the above
argument that Alice cannot communicate with Bob through
entanglement without interaction applies to any two degrees
of freedom that are entangled. Alice's and Bob's measurements
need not be space-like separated events; Bob could make his
measurement to the future of Alice. And the two entangled
degrees of freedom need not be separated in space like the
two particles above; they could be right on top of each
other. To prove this, it is sufficient to note that the
evolution of the reduced density matrix $\rho_2$ of particle
$2$ is governed entirely by the Hamiltonian of particle $2$
because there is no interaction between particles $1$ and
$2$. Therefore, whatever measurement Alice makes on particle
$1$ would affect $\rho_1$ but not $\rho_2$. Hence, the
outcomes of Bob's measurements on particle $2$ that are
determined by $\rho_2$ are unaffected by Alice's
measurements.

Thus Alice's inability to send a signal from event $a$ to
event $b$ by means of entanglement between two non-
interacting systems is independent of whether $a$ and $b$ are
space-like, time-like or null separated. However, Alice may
send a signal using the time-evolution of the wave function
when $a$ and $b$ are time-like or null separated. For
example, Alice and Bob may agree beforehand to measure a particular component of
spin, say $S_z$, and that if Alice
sends a spin-up (spin-down) particle to Bob then Alice means
yes (no). Suppose Alice wishes to send the signal `yes'. She
then measures a spin component on the wave function $\psi$ of
a spin-half particle at time $0$ in the neighborhood of $a$.
If the outcome is spin-up she does nothing, but if the
outcome is spin-down she rotates it to make
it spin-up or she keeps measuring other spin-half particles
until she gets one in the spin-up state. Then she sends
$\psi(0)= \psi_\uparrow$ to Bob. During the subsequent time
evolution, there is rotational symmetry. Therefore, due to
conservation of angular momentum, at time $t$, $\psi(t)$ will
have spin up. Bob then measures $S_z$ in
the neighborhood of $b$ that is in the  future of $a$. He
finds it be in the state $\psi_\uparrow$ with $100\%$
proabability, and feels elated because Alice has said `yes'.

However, if Alice had not communicated to Bob beforehand
which spin-component she will measure, Bob cannot determine
the signal she had sent. This is because the outcome of Bob's
measurement of the spin-component in a general direction is
indeterminate. Alice then would have to send a large number
of particles in the state $\psi(0)$ at time $t=0$ to Bob so
that Bob may statistically determine by means of his own
measurements the signal that she sent. Even this would
not be possible if Alice is not allowed to get around the
indeterminacy of her measurements and put her particle(s) in
the state $\psi(0)= \psi_\uparrow$ as in the first case of trying
to communicate via entanglement mentioned above. Hence {\it the
metaphysical connection, due to entanglement, between
$\psi_\uparrow$ and $\psi_\downarrow$ is similar to
the metaphysical connection, due to Schr\"odinger evolution,
between $\psi(0)$ and $\psi(t)$, because of the indeterminacy
in the outcomes of measurements} of Alice and Bob. But, as seen above,
there is nevertheless an important distinction between the two connections or
correlations, which is called physical causality in the next section.

\section{Physical and Metaphysical Causalities}
\label{causality}

It is necessary to distinguish between two types of ``causality''
at this point. By metaphysical or deterministic causality will be
meant the relation between occurrences $\alpha$ and $\beta$ that
exists if the occurrence $\alpha$ always produces, determines or
necessitates the occurrence $\beta$. But the ability to
communicate information, albeit probabilistically, is here called
physical causality. More precisely, physical or probabilistic
causality is the relation between the occurrence $\alpha$ and
occurrences $\beta_1, \beta_2,\beta_3,....$ that holds if given
the occurrence of $\alpha$ we can predict the probabilities of
$\beta_1, \beta_2,\beta_3,....$. In the examples in section
\ref{indeterminism}, if Alice and Bob do not measure the same
component of spin, say by prior agreement, then Alice's
measurement, call it $\alpha$, does not necessitate the result
$\beta$ of Bob's measurement, whether or not these two
measurements are separated by a space-like, time-like or null
intervals. So, there is no metaphysical causality in all these
cases. However, Alice can influence the {\it probability} of the
outcome of Bob's measurements if Bob makes his measurements to the
future of Alice, as in the example in the previous section in
which she sends him polarized particles, but not when the two sets
of measurements are space-like separated. This is an example of
physical causality. In this section it will be argued that
symmetries are responsible for some common statements involving
physical causality.

One such statement is Einstein causality according to which there
can be {\it physical causality} between time-like or null
separated events but not between space-like separated events. As
mentioned above, this is realized only probabilistically in
quantum phenomena. But Einstein himself was deeply attached to
metaphysical causality, as shown from his following statement
\cite{be1971}:

\small ``... I should not want to be forced into abandoning strict
causality without defending it more strongly than I have so far. I
find the idea quite intolerable that an electron exposed to
radiation should choose of its own free will, not only its moment
to jump off, but also its direction. In that case I would rather
be a cobbler, or even an employee in a gaming-house, than a
physicist.'' \normalsize\\It is ironical that Einstein who
overthrew three of the four metaphysical relations mentioned in
section \ref{metaphysical}, could never give up the fourth
metaphysical relation of causality.

The only way to experimentally test physical causality is by
means of a large number of trials of the form $
(\alpha,\beta_1), (\alpha,\beta_2), (\alpha, \beta_3),....$. The
relative frequency of each distinct pair in this experiment,
as the number of trials tend to infinity, is the probability
of occurrence of this pair. However, in a given trial, say
$(\alpha, \beta_3)$, we may ask, as Einstein does implicitly
in the above statement, why is it that $\alpha$ was followed
by $\beta_3$ and not $\beta_1, \beta_2$ or $\beta_4,..$? This
absence of metaphysical causality, due to the indeterminism
in quantum phenomena, should make us re-examine the meaning
and validity of physical causality as well. Suppose Alice
sends a large number of spin-half particles in state
$\alpha$, say the spin-up state, to Bob. By doing experiments
of the above form with a large number of trials, Bob
determines to a very high probability the state $\alpha$.
This is possible in this instance because of conservation of
angular momentum that ensures that all the particles are in
the same spin state when they reach Bob. But this
conservation of angular momentum is due to rotational
symmetry.

In general, conservation laws are due to symmetries. And if
the state is not an eigenstate of the conserved quantity then
the conservation is realized only statistically, i.e. the
expectation value of the conserved quantity is preserved in
time. The equality of the expectation values of the conserved
quantity at two different times is therefore like physical causality,
because of the
probabilistic manner in which they are both determined,
and does not imply metaphysical
causality. But
the symmetries are not probabilistic as far as we know.
%Since causality is associated with dynamical laws, which give
%conservation laws,
This suggests that symmetries may be more
basic than dynamical laws that may actually be effective laws
arising from symmetries.

It is often stated that Einstein causality is incorporated in
quantum field theory by the requirement that field operators
at events that are space-like separated are independent in
the sense that they must commute if they are Bosonic fields
and anti-commute if they are Fermionic fields. I.e. if
$\phi_m(x)$ are the components of the various fields in the
theory, where $x$ stands for the space-time coordinate $({\bf
x},t)$, then
\begin{equation}
[\phi_m(x),\phi_n(y)]_{\pm} =0
\label{independence}
\end{equation}
whenever $x$ and $y$ are space-like separated. The  $\pm$ in
(\ref{independence}) refers to anti-commutator if the fields
are Fermionic and commutator if the fields are Bosonic.

However, (\ref{independence}) does not require Einstein
causality and may be introduced in order that a local quantum field
theory is Lorentz invariant \cite{we1995}. To see this,
consider the time evolution operator $U$ for quantum states
in the interaction picture of a canonically
quantized field theory that is generated by the interaction
Hamiltonian $H_I(t)$.
Since $H_I(t)$ represents the interaction
energy, and the locality assumption requires that this energy
is the sum of energies in all the different regions of space,
$$H_I(t)=\int d^3x {\cal H}(x),$$
where $ {\cal H}(x)$ is the interaction Hamiltonian density.
It follows that

$$U=T\exp(-{i\over\hbar}\int H_I(t) dt)= 1+\sum_{n=1}^\infty
{1\over n!} (-{i\over\hbar})^n\int d^4x_1 d^4x_2 .... d^4x_n
~T\{{\cal H}(x_1) {\cal H}(x_2)
....{\cal H}(x_n)\} $$
where $T$ denotes the time ordering operation meaning that it
orders the operators from left to right in the monotonically
decreasing order of the values of their argument $t$. This
ordering is independent of the chosen Lorentz frame if every
pair ${\cal H}(x_r), {\cal H}(x_s) $ commute whenever $x_r,
x_s$ are space-like separated. This is guaranteed by the
independence  of fields at space-like separated events given
by (\ref{independence}).
Thus `Einstein causality' in a local quantum field theory may be
regarded as due to the Lorentz group of symmetries. This is
analogous to how the same symmetries discarded the
metaphysical relations (2) and (3) of Newtonian absolute simultaneity and
causality between simultaneous events during the
creation of special relativity, as mentioned in section \ref{metaphysical}.

\section{Objective Probabilities and Probability Amplitudes}
\label{probability}

The metaphysical causality, mentioned above, was associated
with the metaphysical necessity of causal dynamical laws
\cite{an1999}. Therefore, since this metaphysical causality
was discarded above, it is no
longer necessary to assume the
metaphysical necessity that is responsible for causal laws.
An advantage of discarding metaphysical causality is that
the indeterminism of quantum phenomena may then be deduced,
instead of postulating it {\it a priori}. This solves the statistical aspect of
the measurement problem,
mentioned in Einstein's statement quoted in section \ref{causality}.
The only causality which now remains is the physical causality
that is determined by the operator $U$, and which is realized
only probabilistically.

The above arguments suggest, however,
that symmetries are more fundamental than physical causality.
I shall therefore, from now on, assume A) {\it there are no causal
dynamical laws}. This implies that physical processes cannot
be deterministic, because there is nothing compelling a
physical system to evolve in a definite manner. It follows
that we can only assign probabilities for physical processes,
which is consistent with the experimentally observed
intrinsic indeterminism of quantum phenomena, as mentioned above.
I shall assume
also that B) {\it the probabilities of physical processes are
invariant under a group of symmetries}.

According to classical physics, a particle goes from an event $a$
to an event $b$ along a definite path that is determined by the
laws of classical physics. If we discard causal dynamical laws, in
accordance with assumption (A), then the particle need not take a
definite path and all paths between $a$ and $b$ are equally
probable. If we give up only determinism which is associated with
metaphysical causality, then it is possible for the probabilities
for the paths to be different, e.g. the classical path may have
probability greater than all the other paths. But this would be a
causal law that is probabilistic. I am taking assumption (A),
however, to imply a maximal violation of the classical causal
dynamical law, which is why all paths are given equal probability.
Physically, this means that if we observe precisely whether the
particle takes an {\it arbitrary} path between $a$ and $b$ then it
would take this path with $100\%$ probability; this makes all
paths between $a$ and $b$ have the same probability\cite{ah1980}.
This condition will be discussed more later. More generally, the
paths may be in the configuration space of a more complicated
physical system. The preference for configuration space instead
of, say, the momentum space of the system is because the
observation of the path by an apparatus needs to be realized by an
interaction between the apparatus and the system, and all
interactions are local at presently accessible energy scales. In
order for such an observation to be like any other physical
process, reality needs to be defined relationally as the outcome
of interactions, which will be done in section \ref{measurement}.
For simplicity and ease of visualization, I shall continue to
treat the possible paths of the system as space-time paths of a
particle.

If we suppose naively that the probability of the particle to go
from $a$ to $b$, denoted $P(b,a)$, is the sum of the
probabilities of the possible individual paths then, since there
are an infinite number of equally probable paths, the probability
of each path is zero. We may try to mathematically implement this
equal probability rule by defining on this set of paths a measure
and a probability density that is positive and constant. Then if
the volume of this set of paths with respect to this measure is
infinite, the integral of this probability density would also be
infinite and cannot equal $P(b,a)\le 1$. Restricting the volume
of the infinite dimensional manifold of paths to be finite would
be artificial and would amount to introducing a law. Therefore,
there must be cancellation between different paths in order to
obtain a sensible result for $P(b,a)$. This may be achieved by
introducing the {\it probability amplitude} for each path such
that the set $\cal A$ of probability amplitudes is a group under
the operation of addition. If we require that the probability
amplitudes should first be added before forming the probability
from this sum, then adding these amplitudes would give the
required cancellation. Also, the probability of a system going
from $a$ to $c$ through $b$ should be the product of its
probabilities to go from $a$ to $b$ and $b$ to $c$. This suggests
that it should be possible to multiply probability amplitudes.
Then $\cal A$ should form an algebra. To obtain a probability
that is a non-negative number from the probability amplitude, one
may introduce a norm, denoted by $||$, on $\cal A$. The above
mentioned multiplicative property of probabilities suggests that
$|\psi\phi|=|\psi||\phi|$, for probability amplitudes $\psi$ and
$\phi$. A theorem by Hurwitz \cite{hu1898}, generalized by Albert
\cite{al1947}, states that $\cal A$ should then be the reals,
complex numbers, quaternions or octonions \cite{ad1995}.

Octonions may be excluded as candidates for probability
amplitudes due to the non associativity under multiplication, if
addition is also taken into account. Consider the case of an
electron that can either go through two screens with a single
slit in each or go around both screens to reach the point where
it is detected. If we use octonions as probability amplitudes,
then the probability of detection is determined by the norm $|
(\psi_1\psi_2)\psi_3+\phi|$ or $|\psi_1(\psi_2\psi_3)+\phi|$
where the octonions $(\psi_1\psi_2)\psi_3$ or
$\psi_1(\psi_2\psi_3)$ are the amplitudes to go through the two
slits and the octonian $\phi$ is the amplitude to go around the
two slits. But the above two norms are unequal in general. For a
fixed set of amplitudes, one could try to remove this ambiguity
by choosing a particular order of multiplication, e.g.
$((\psi_1\psi_2)\psi_3)...$ which violates time reversal
symmetry. Also, if we divide a path into segments, then it is not
possible to see how one could define a product of amplitudes
associated with the segments that would be independent of the
choice of this division.

For the probabilities of different paths between a pair of events
to be equal, in accordance with assumption (A), it is reasonable
for the norms of the corresponding probability amplitudes to be
equal. If $\cal A$ is taken to be the field of real numbers then
these amplitudes can only differ in sign, and the sum of these
infinite number of amplitudes would not give a sensible result. It
is perfectly possible {\it a priori} to have a physical theory
that uses only real numbers, e.g. classical physics. But such a
theory would not satisfy assumption (A), above, if we allow for
infinite number of ways in which a system may go from a
configuration $a$ to a configuration $b$ with equal probabilities.
The probability amplitude should therefore be taken to be a
complex number or quaternion. Adler \cite{ad1995} has found it not
possible to construct a path integral using quaternions. We
therefore take $\cal A$ to be the field of complex numbers,
because the present approach leads naturally to a path integral,
as will be seen explicitly in the next section.

\section{Origin of the Born Rule and Quantum Dynamics}
\label{born}

The question arises as to how the probability $P(b,a)$ may be
obtained from the sum of probability amplitudes, denoted $K(b,a)$.
To answer this, consider the double slit experiment and, for
simplicity, suppose that there are just two paths, $\gamma_1$ and
$\gamma_2$ for a particle to go from an event $a$ at the source to
an event $b$ at the screen through slit $1$ and slit $2$,
respectively. Then $ K(b,a)= \psi_1 +\psi_2$ where $\psi_1$ and
$\psi_2$ are the complex probability amplitudes for the paths
$\gamma_1$ and $\gamma_2$, respectively. Let $\tilde P(\psi)$
denote the probability of the probability amplitude $\psi$.

Now if we observe through which slit the particle went through
then the probability is the sum of the probabilities of going
through either slit. This is a consequence of the relational
reality which will be introduced in the next section according to
which the interaction of the particle with another system
determines the state of each of them relative to the other.
Interaction, by definition, changes the phase of the probability
amplitude. Since a device that interacts with the particle to
determine whether it took path $\gamma_1$ is also quantum
mechanical, the phase of $\psi_1$ is uncertain\footnote{The usual
description of decoherence regards it as occurring due to the
entanglement of the system with states of the environment that are
orthogonal due to the influence of the system. But it has been
shown that \cite{fe1963}, equivalently, decoherence may be
regarded as due to an uncertain phase  difference $\theta$ between
interfering probability amplitudes due to the influence of the
quantum environment. Specifically, if the average
$<e^{i\theta}>=0$, then there is decoherence.}. I shall suppose
that this interaction is such that the phase of $\psi_1$ is
completely uncertain. Averaging over this uncertain phase should
give the sum of the probabilities for the two paths:
\begin{equation}
{1\over 2\pi}\int_0^{2\pi} d\theta_1 ~\tilde P(\psi_1+\psi_2)=
\tilde P(\psi_1) + \tilde P(\psi_2) \label{whichway}
\end{equation}
where $ \tilde P(\psi_1)$ and $\tilde P(\psi_2)$ are the
probabilities for the paths $\gamma_1$ and $\gamma_2$, and
$\theta_1=\arg \psi_1$, $\theta_2=\arg \psi_2$. This averaging may
be physically visualized by means of an ensemble of particles that
pass through the double slit and strike the screen as follows. The
particles acquire all possible relative phases for $\psi_1$ and
$\psi_2$ with equal probabilities. Each relative phase gives a
corresponding probability distribution for detecting the particle
on the screen. Then (\ref{whichway}) is the average of this
probability distributions.

Since $\theta_1$ in the left hand side of (\ref{whichway}) is
integrated over, $\tilde P(\psi_1)$ in the 
right hand side is independent of
$\theta_1$. Hence,
$\tilde P(\psi)$ is a function of $|\psi|$ only. Since $\tilde P(\psi)$
is a non-negative function, it is
reasonable to suppose that $\tilde P(\psi)=|\psi|^n$, where $n$ is
a non-negative integer. Suppose also that $ \tilde
P(\psi_1)=\tilde P(\psi_2)$, which is the case if $\gamma_1$ and
$\gamma_2$ are the only paths available to the particle. Then,
$|\psi_1|=|\psi_2|=a$, say, so that
$|\psi_1+\psi_2|=2a|\cos(\theta/2)|$, where
$\theta=\theta_1-\theta_2$. Therefore, (\ref{whichway}) reads
\begin{equation}
{2^n\over 2\pi} \int_0^{\pi} d\theta |\cos^n\theta|= 1.
\label{thisway}
\end{equation}
Now for any non-negative integer $m$,
$$ \int_0^{\pi} d\theta |\cos^{2m+1}\theta|
=2\int_0^{\pi/2} d\theta \cos^{2m+1}\theta ={2^{2m+1}(m!)^2\over
(2m+1)!}$$ Since this is a rational number, (\ref{thisway}) cannot
be satisfied by odd $n$. Therefore, $n=2m$. Now,
$$ \int_0^{\pi} d\theta \cos^{2m}\theta = {(2m)!\over 2^{2m}(m!)^2}\pi$$
Hence, (\ref{thisway}) reads
\begin{equation}
{(2m)!\over 2(m!)^2}=1
\label{condition}
\end{equation}
This is satisfied for $m=1$. But for $m>1$, it is easily shown
that ${(2m)!\over 2(m!)^2}>1$ and therefore (\ref{condition}) is
not satisfied. Also, for $n=0$ clearly (\ref{thisway}) is not
satisfied. Hence, {\it (\ref{whichway}) is satisfied with $\tilde
P(\psi)=|\psi|^n$ if and only if $n=2$}, i.e.
\begin{equation}
\tilde P(\psi)=|\psi|^2
\end{equation}
Hence, $P(b,a)= |K(b,a)|^2$, which gives the Born rule. This of
course is in agreement with the observed interference pattern that
results if we do not observe which slit the particle went through.

Since
all paths are equally probable, in accordance with assumption
(A), it follows that the probability amplitude assigned to an
arbitrary path $\gamma$ joining $a$ and $b$ should be
$N\exp[i{\cal S}(\gamma)]$, where $N$ is independent of $\gamma$
and $\cal S$ is real. Hence,
\begin{equation}
K(b,a) = \sum_{\gamma} N\exp[i{\cal S}(\gamma)],
\label{pi}
\end{equation}
Assumption (B) then implies that the probability $|K(b,a)|^2$ is
invariant under the symmetry group. A sufficient condition for
this is that $\cal S$ is invariant under the symmetry group. The
highly successful Feynman path integral formulation of quantum
mechanics assumes (\ref{pi}) with ${\cal S}(\gamma)=S_\gamma
/\hbar$, where $S_\gamma$ is the classical action for the path
$\gamma$. Under this assumption it is easy to understand why the
classical limit corresponds to the particle taking the trajectory
for which $S$ is an extremum: In the classical limit $S_\gamma
>>\hbar$ for all possible trajectories, but the amplitudes for
trajectories far away  from the classical or extremal trajectory
cancel out each other, while those in the neighborhood of the
extremal trajectory add constructively.

As mentioned earlier, quantum mechanics maximally violates the
laws of classical physics because the classical laws constrain the
particle to move along the classical trajectory for which the
classical action is an extremum, whereas quantum mechanics gives
equal probability to all possible trajectories. But if the action
in (\ref{pi}) is confined to be the classical action, as in
ordinary quantum mechanics, then this would constitute a law,
albeit probabilistic. Therefore, in accordance with the principles
(A) and (B) above, I require a maximal violation of the laws of
quantum mechanics as formulated by Feynman by postulating the
following hypothesis: $S_\gamma$ {\it contains all terms that are
invariant under the symmetries}. For this hypothesis to be in
agreement with observation, it is of course necessary that the
resulting quantum theory should be approximately in agreement with
quantum mechanics that uses only the classical action for all the
experiments that we have performed so far.

In order to do this, we need to replace $S_\gamma$ by an action
$S$ made from all the fields which are obtained from
representations of the symmetry group. The development of
effective field theories \cite{effective} have enabled the
inclusion into $S$ all the terms that are allowed by the
symmetries that can be formed from the fields in the standard
model. Only the lowest order terms are conventionally
renormalizable, in the sense that they require a finite number of
counter terms to cancel the ultraviolet divergences, which may
thus be included into a finite number of coupling constants. This
makes the standard model unique apart from the values of the
coupling constants which need to be determined by experiments.
While the somewhat unique determination of the standard model by
the requirement of invariance under symmetries and
renormalizability may be an indication of the fundamental role
played by the symmetries, the principles (A) and (B) would
require that we include in $S$ all the terms that are allowed by
symmetries. It turns out that including every term that is
consistent with the symmetries in $S$, as shown by Weinberg
\cite{we1995}, provides counter terms to cancel all the
ultraviolet divergences in the Feynman diagrams. It is the lowest
order terms that we directly observe at presently accessible
energies, which gives the illusion that the action contains only
a finite number of terms, as assumed in the paradigm of laws. If
we go to high enough energies, we should be able to see the other
terms, and measure the associated coupling constants, according
to this hypothesis. The standard model, which was originally
formulated with a finite number of terms in the Lagrangian that
are invariant under the symmetries, defines an effective field
theory, which is sufficient to make all the observable
predictions at the energy scales at which we now do
experiments, according to the present view.

Also, from the above path integral, Schr\"odinger's equation for the fields and
its non relativisitic limit for particles may be obtained in a well known
manner. This may be turned around, and one may start with Schr\"odinger's
formulation and obtain the path integral. If one asks for the probability for
observing the system taking any path between the initial and final states, the
answer turns out to be unity, and the system then acquires the phase associated
with this path that is predicted by Feynman \cite{ah1980}. This result shows the
consistency of Schr\"odinger's or the equivalent Heisenberg formulation of
quantum mechanics with the above assumption that all paths have equal
probability and the result that the phases of the probability amplitudes
associated with the paths at low energies are the Feynman phases.

In the present view, the action is more fundamental
than the `laws' derived by extremizing the action, which
has important physical consequences. It implies that
a given field that obeys an effective law cannot be regarded as
complete unless this law can be obtained from an action
principle, which may require introducing other fields. For
example, the electromagnetic field strength $F_{\mu\nu}$
obeys the Maxwell's equations and the Lorentz force
classically. But to obtain these laws from an action
principle, we need to introduce the potential $A_\mu$. And the
action that is a function of $A_\mu$ then gives rise to new
effects, such as the Aharonov-Bohm effect \cite{ah1959}. While the
Aharonov-Bohm effect may be expressed non locally in terms of
the field strength $F_{\mu\nu}$, its generalizations to
non-Abelian gauge fields cannot be expressed in terms of
only the Yang-Mills field strength $F_{\mu\nu}^i$ even
non-locally, and requires the potential $A_\mu^i$.

The present approach explains the physical causality discussed
in section \ref{causality} in a lawless manner. The probability for any physical
process is obtained from (\ref{pi}), where all possible histories are given
equal probability weights. Therefore,
when Alice's and Bob's measurements are time-like separated, the determination
of the probabilities of the outcomes of Bob's measurements by Alice's
measurement is due to the constraint of the probability amplitudes by the
symmetries, and not because of any causal dynamical law.

\section{The Physical Meaning of Probabilities and the Quantum Measurement
Problem}
\label{measurement}

The probability amplitudes were obtained in the previous section
on the basis that there are no causal laws. The `probabilities'
that are obtained from the probability amplitudes are the
probabilities of `real' events. This raises the question of what
reality means. So far no objective criterion has been provided
for when the probability amplitudes should be converted to
probabilities, which is the quantum measurement problem in the
present language of probability amplitudes (as opposed to wave
functions) \cite{ka1995}.

The notion of reality that has been commonly used is the view that
`existence' is an absolute property or predicate that any
conceivable object does or does not possess. If it has this
property or predicate then it is said to exist, otherwise it is
said not to exist, and this is independent of its interactions
with other objects. I shall call this the notion of absolute
reality.

But from an operational point of view, absolute reality is
meaningless. Consider the statement that there is only one object
in the universe. Whether this object exists or does not exist
cannot be operationally distinguished and seems to be a
meaningless question. However, the existence of two objects may be
given meaning through the interaction between them. This leads to
the notion of {\it relational reality} or interactive reality,
namely that two objects exist in relation to each other if they
interact. According to this view, the state or wave function in
which an object `exists' is meaningless if it is not interacting
with another object. It was mentioned in section
\ref{indeterminism} that, although entanglement between two
objects represents correlations, it is not possible to communicate
by means of this entanglement unless the objects interact; thus
interaction between two objects is necessary for one object to
know the (relative) existence of the other.

Another motivation for introducing relational reality comes from the
criterion of reality formulated sometime ago \cite{an1995}.
According to this if two objects interact in such a way so as
to satisfy the action-reaction principle, then both objects exist.
This criterion was then used as a sufficient condition for
establishing the reality of objects in particular cases.
But the relational reality introduced above is symmetrical
with respect to the two objects, and may provide a deep reason
why the action-reaction principle is {\it always} satisfied in
nature, or this may conversely be regarded as indication of
relational reality. Indeed, I shall use the action-reaction
principle here as a necessary and sufficient condition for
reality. As a particular application, according to the
De Broglie-Bohm hypothesis, the wave guides the particle
to move along a particular trajectory, but the particle
itself does not react back on the wave \cite{bo1952}.
According to the above symmetric criterion of reality
therefore we cannot regard the particle being
in this trajectory as real.

The above relational reality removes some of the paradoxes
associated with the quantum measurement problem. For example, the
Schr\"odinger cat may be inside a box in the alive state
$|\psi_a>$ relative to the interactions it undergoes with itself
and the box. But an observer outside the box, who so far has not
interacted with the cat, could in principle observe the cat in the
state $|\psi_c>={1\over\sqrt 2}(|\psi_a>+|\psi_d>)$ with
probability 1, where $|\psi_d>$ is the state of the dead cat, {\it
provided} there is an interaction between her or her apparatus and
the cat such that $|\psi_c>$ is an eigenstate of the interaction
Hamiltonian $H_I$. There would be no contradiction between the two
states of the cat, namely $|\psi_a>$ and $|\psi_c>$ because these
are states that the cat has relative to the interactions it
undergoes with two different systems. But in practice, there is no
interaction for which the cat would have the state $|\psi_c>$,
because if there were one, then $<\psi_a|H_I|\psi_d>\ne 0$, which
is not possible because of the large number of degrees of freedom
of the cat. Therefore, it is not possible to give relational
reality to the superposition $|\psi_c>$. Also, the outside
observer is interacting with the cat through the gravitational
field and via the box with which they both interact; so they form
a single interactional reality. For example, she could verify in
principle that the cat is alive by interacting with its time
dependent gravitational field outside the box as it moves inside
the box, without opening the box. Hence, when she opens the box,
she will see the cat in the alive state. In general, the relational reality of the cat is that it is alive or dead.

But for microsystems, it is easy to produce examples like the one
above. For example, consider a double slit experiment with
electrons. The state of each electron when it passes the double
slilt screen is ${1\over \sqrt2}(|\psi_1>+|\psi_2>)$ relative to
the screen because of its interaction with the screen. Suppose
that near one slit a neutron is introduced which is in a spin-up
state $|\uparrow>$, with respect to the $z-$axis of a Cartesian
coordinate system, which is verified by a separate interaction of
an apparatus with the neutron. If the neutron does not interact
with the electron, then due to conservation of angular momentum
which in turn is due to rotational symmetry, the neutron may
subsequently observed by means of a suitable interaction to be in
the same state with probability $1$. The arrangement is such that
whenever the neutron interacts with the electron, it undergoes a
spin flip, i.e. its state is $|\downarrow>$ relative to the
electron, while the state of electron relative to the neutron is
then $|\psi_1>$. But an external observer may observe the combined
system of the neutron and electron in the state $$|\psi>={1\over
\sqrt2}( |\psi_1>|\downarrow> +|\psi_2>|\uparrow>)$$ with
probability $1$, relative to an interaction that she or her
apparatus has for which $|\psi>$ is an eigenstate of $H_I$. On the
other hand, the neutron and the electron `observe' each other,
whenever they interact, in the state $|\psi_1> |\downarrow> $ with
probability $1/2$. Of course, neither particle can express what
they `observe,' unlike human beings. But this statement is not
empty because it may be verified by observing the state of the
neutron to be $|\downarrow>$ in half of an ensemble of such
experiments.

The third example that I shall consider is the protective
observation of the extended wave function of a single particle
\cite{ah1993}. The key to the protective observation is that the
combined system of the particle being observed and the particle
that does the probing is in an {\it unentangled} state $\psi\phi$,
where $\psi$ and $\phi$ are the wave functions of the two
particles, respectively. This is achieved by having $\psi$
originally in an eigenstate of the Hamiltonian and making the
interaction between the two particles adiabatic and weak so that
$\psi\phi$ continues to evolve as an approximate eigenstate of the
Hamiltonian, according to the adiabatic theorem. For example,
$\psi$ may be the ground state of a proton inside a box and $\phi$
may be the localized wave packet of an electron slowly moving in the box.
By sending several such electrons, from the behavior of the
corresponding states $\phi$, it is possible in principle to
reconstruct $\psi$, even though this is the extended wave function
of a single particle. The usual non adiabatic strong measurements,
however, would enable the reconstruction of an extended $\psi$
only statistically by putting sequentially many protons inside the
box in the state $\psi$, and observing each of them with electrons.
This is because in the usual measurements
there will be an entanglement between the two systems, which
prevents us from saying which state the proton is in. If we try to
find this by means of a subsequent measurement on the electron,
the entangled wave function of the combined system appears to
undergo a sudden unpredictable change,
which brings up the measurement problem
in the language of wave functions.

There is now the following paradox: Suppose we observe $\psi$
protectively (adiabatically) and conclude that it is real
or ontological. And then we do the usual measurement,
which leads to the ``collapse'' of this wave function to a
localized state $\psi'$. For the new measurement, $\psi$ is only used
epistemologically to predict the probability of the new state $\psi'$ by
means of the Born rule. So, what happened to the reality of $\psi$? This
paradox is resolved if we give up the notion of absolute reality, and
accept that both $\psi$ and $\psi'$ are real only in relation to the
interaction that they undergo with the systems they interact with. The
sudden change from $\psi$ to $\psi'$ is not paradoxical if we do not
assign absolute reality to either wave function, but instead assign
relational reality to them, which requires that such a change should take
place because the interaction that determines this relational reality has
changed.

In all three examples, above, in the assigment of relational
reality to one of two interacting systems, the following
condition is satisfied, which will be called $R$: {\it Two
systems interact with each other in such a way that their states
remain unentangled, and the interaction satisfies the
action-reaction principle.} This suggests using the condition $R$
as necessary and sufficient  criterion for the relational reality
of states or wave functions, in general. I shall therefore assume
this to be the case, and postulate that the quantum probabilities
of the previous section apply to states that satisfy the
condition $R$.

The influence of each state on the other provides information
about the former via the interaction, when $R$ is satisfied. To
illustrate this, consider the commonly encountered situation in
which the Hamiltonian for two systems $S_1$ and $S_2$ is
$H=H_1+H_2+H_{12}$. The subscripts $1$ and $2$ designate
operators that act on the Hilbert spaces of $S_1$ and $S_2$,
respectively, and $ H_{12}$ is the interaction Hamiltonian.
Suppose now that $ H_{12}=g(t)A_1A_2$, where $g(t)$ is a c-number
coupling constant that represents the turning on and off of the
interaction. In the impulse approximation, $g(t)$ is very large
for a short period of time so that the free Hamiltonian is
negligible compared to the interaction Hamiltonian during this
period. Then, if the combined system $S_1+S_2$ was initially in
the unentangled state $\psi_1\psi_2$, where $\psi_1$ is an
eigenstate of $A_1$ with eigenvalue $a$, $S_1+S_2$ will continue
to be unentangled during the interaction. But the response of
$S_2$ to the interaction, which may be determined by subsequent
observations (interactions), will depend on the eigenvalue $a$ of
$A_1$. This is the usual von Neumann measurement specialized to
the case in which $S_1$ was already in an eigenstate of the
observable $A_1$ at the beginning of the interaction. On the
other hand, more generally, if $\psi_1$ were not an eigenstate of
$A_1$ then the state of $S_1+S_2$ that is obtained by solving
Schr\"odinger's equation is an entangled state $\psi_{12}$ due to
the interaction. According to the present view, $\psi_{12}$
should be regarded as the probability amplitude from which the
probabilities for the various states in this superposition that
satisfy condition $R$, such as the above $\psi_1\psi_2$, may be
obtained. The only way to give relational reality to $\psi_{12}$
is to let it interact with another system with state $\phi$ such
that $\psi_{12}\phi$ satisfies condition $R$ which makes
$\psi_{12}$ real with respect to $\phi$.

We may now take the opposite limit in which $g(t)$ is small and
varies slowly or adiabatically, so that the interaction
Hamiltonian may be regarded as a perturbation of the free
Hamiltonian. Also, suppose that $\psi_1$ is a non degenerate
eigenstate of $H_1$ and $g(t)$ does not contain any frequency
component that connects $\psi_1$ and the nearest eigenstate. This
is the case of protective measurement \cite{ah1993} in which  the
combined system remains in the unentangled state; but the
response of $\psi_2$ depends on $<\psi_1|A_1|\psi_1>$, which
generalizes the eigenvalue of $A_1$. The third example is the
weak measurement of Aharonov (see for example \cite{ah1988}) in
which $H_{12}$ is so weak that it causes only a small
entanglement to the initial or preselected unentangled state of the two systems.
One then makes a postselection of a particular state $\psi_1$ of system $S_1$.
This puts the combined system in an unentangled state. From the state of system
$S_2$, compared to its initial state, the `weak value' $ A_{1w}=
<\psi_1|A_1|\phi_1> /<\psi_1|\phi_1> $ is obtained, where $\phi_1$ is the
preselected state of $S_1$. The preselection of $\phi_1$ and the postselection
of $\psi_1$ mean that each of these states has relational reality with respect
to another system that interacts with it satisfying condition $R$ at the times
of preselection and postselection, respectively. Hence, the information obtained
about these states, $\phi_1$ and $\psi_1$, from $ A_{1w}$ is consistent with the
relational reality assigned to these states.

In practice, we obtain the information about the
state $\psi_1$ through its interaction with $\psi_2$ by choosing
$S_2$ to be macroscopic or from a subsequent interaction of $S_2$
with a macroscopic system. But this is only because we are
macroscopic beings, and it does not imply any necessary asymmetry
between the two interacting systems, objectively speaking. The
above relational reality condition $R$ is explicitly symmetric.
For example, in the Stern-Gerlach experiment with neutrons, a
given neutron is in the spin-up or spin-down state relative to the
inhomogeneous magnetic field of the magnet, and the magnet
relative to the neutron is then in one of two possible states
that have opposite momenta during their interaction. This is
because each of these unentangled states of the combined system
satisfies $R$, and the above objective probabilities require that
only one of these possibilities are realized.  The magnet may be
microscopic or macroscopic. If it is microscopic, then it becomes
necessary {\it for us} to verify the response of the magnet by
means of its interaction with a macroscopic system.

The nonentanglement stipulated in condition $R$ may be satisfied
only approximately in practice. But if one views reality as fluid
and not concrete then this is not a problem. The notion of stark,
concrete, absolute reality that we use in our daily lives,
according to which something {\it is} or {\it is not}, may be
useful for our survival, for example, if we encounter a tiger in
a jungle. For this reason, our brains may have acquired the
notion of absolute reality during evolution by natural selection.
This may be due to the stability of certain unentangled states in
our brains due to the nature of the interactions between them.
But an objective examination of the interactions between
microsystems suggests that absolute reality is not valid. The
symmetrical relational reality being postulated here, instead,
does {\it not} lead to solipsism because each of us is
interacting by means of the gravitational and electromagnetic
fields with all the other objects, which are therefore real to
us, and we are to them. But the state in which we observe another
system depends on the nature of our interaction with it.

\section{Discussion and Speculation}

According to the present view, an arbitrary state of any isolated
object represents only the potentialities for the relational
reality that it may have if and when it interacts with another
object. An `event' for which quantum mechanics assigns a
probability of occurrence is the interaction of a pair of states
of two systems that satisfy condition $R$. Since the interactions
are local (due to the fundamental interactions of gravity and
gauge fields being local), the two states must be localized in
order for them to remain unentangled as required by $R$. This
explains why the world of events that we observe appears to be
classical even though quantum mechanics is universally valid, in
the present approach. It also appears to explain the origin through these events of the
space-time description, which is different from the Hilbert space
description of quantum mechanics.

The interpretation of quantum mechanics which emerges from the
above analysis is intermediate between the Copenhagen and the Everett
interpretations. It may be reached from the Everett interpretation
by replacing the notion of absolute reality in this
interpretation, which makes all the worlds
that are nearly orthogonal because of
decoherence to be real, by relational reality that makes only one
of these Everett worlds to have relational reality. The reason for the latter
conclusion
is that there is no interaction that connects two Everett worlds,
as in the case of the alive and dead states of the cat mentioned
above, i.e. a superposition of the states of two Everett worlds cannot interact
with another object so that condition $R$ is satisfied. Therefore, a
superposition of two Everett worlds cannot have
reality with respect to another object.
Furthermore, the {\it
objective probabilities}, introduced in sections \ref{probability} and \ref{born},
give a probabilistic prediction for the realization of one of these
Everett worlds at each instant of time.
On the other hand, the Everett picture is
deterministic and therefore cannot contain objective
probabilities; hence probabilities may be introduced only by
coarse-graining, i.e. through ignorance of the precise initial
conditions, as in classical statistical mechanics where also there
are no objective probabilities. It is not clear how the
empirically successful Born rule that uses the geometry of Hilbert
space could be obtained from coarse-graining. But in the present
interpretation, these probabilities emerge
naturally and objectively from the
denial of fundamental causal laws, and the requirement of
invariance under symmetries, as shown in sections \ref{probability} and \ref{born}.

The present interpretation may also be reached from the Copenhagen
interpretation if in addition to the denial of absolute reality in
the latter interpretation, we introduce relational reality. This
removes the anthropomorphic concepts from the Copenhagen
interpretation, such as the need for a `classical' measuring
apparatus with a human observing it, or introducing `measurement'
by a human or at least a macroscopic system as a special
interaction. As Niels Bohr said, we need to specify the entire
experimental arrangement before assigning reality to any part of
it. But this is necessary only because relational reality is
determined by all the interactions in the entire experimental
arrangement. It is not necessary for the specified experimental
arrangement to contain a macroscopic subsystem. The present
interpretation therefore abolishes the notion of `measurement' as
a special interaction, and provides an objective, but relational,
description of this process assuming the universal validity of
quantum mechanics.

If absolute reality is replaced by relational reality then it
would seem that any universe that is made of interacting parts and
is mathematically consistent may have relational reality. I shall
call the collection of such hypothetical universes the {\it
polyverse}. This speculation, because it is based on the notion of
reality being purely relational, provides an answer to the old
philosophical question of why there is something and not nothing.
Each universe contained in the polyverse has no reality as a whole
because none is interacting with any other. However, parts of each
universe have relational reality due to their mutual interaction.
Incidentally, this implies that the wave function of the universe
as a whole has no reality, but it is meaningful to assign a wave
function to any part of the universe in relation to another part
with which it is interacting. Although none of the other universes
in the polyverse has reality to us, the polyverse nevertheless has
a physical consequence in explaining why the symmetries and
coupling constants of our universe are such that they allow life
to evolve, without assuming that our universe was a special
creation to allow this \cite{an1999}. In other words, the
polyverse provides a physical basis for an anthropic principle. If
we assume, on the other hand, that the polyverse consists of just
one universe, then the only alternative to the special creation
hypothesis is to try to discover a principle that explains the
symmetry groups and values of the coupling constants in our
universe.

\newpage
\noindent{\bf Acknowledgments}
\bigskip

I thank Yakir Aharonov and John A. Wheeler for valuable and
inspiring discussions over the years. I also thank Stephen L.
Adler, Harvey Brown, Todd Brun, Daniel Dix, Lucien Hardy, Ralph Howard,
Parameswaran Nair, Roger Penrose, Anthony Short, and
Leo Stodolsky for informative and stimulating discussions. Daniel Dix, Ian Percival, Maxim
Tsypin and William Wootters are thanked for their comments on an
earlier version of this paper. This research was partially
supported by a NSF grant and a Fulbright Distinguished Scholar Award.


\begin{thebibliography}{99}


\bibitem{an1999}
J. Anandan, Foundations of Physics, {\bf 29,} no. 11, 1647-
1672 (1999), quant-ph/9808045.

\bibitem{fe1985}
R. P. Feynman (1985) {\it The Strange Theory of Light and
Matter} (Princeton Univ. Press), chapter 1.

\bibitem{ei1935}
A. Einstein, B. Podolsky and N. Rosen, Phys. Rev. {\bf 47,} 777 (1935).

\bibitem{fe1948}
R.P. Feynman, Rev. Mod. Phys. {\bf 20,} 367-87 (1948); R. P.Feynman
and A. R. Hibbs {\it Quantum Mechanics and Path
Integrals,} (McGraw-Hill, New York, 1965).

\bibitem{bo1952}
D. Bohm, Phys. Rev.,
{\bf 85,} 166-193 (1952).

\bibitem{ev1957}
H. Everett, Rev. Mod. Phys. {\bf 29,} 454-462 (1957).

\bibitem{be1971}
A. Einstein and M. Born, The Born-Einstein Letters, 29 April 1924.

\bibitem{we1995}
Steven Weinberg, {\it The Quantum Theory of Fields I} (Cambridge
University Press, Cambridge 1995), pp.144, 145, 198.

\bibitem{ah1959}
Y. Aharonov and D. Bohm, Phys. Rev. {\bf 115,} 485 (1959).

\bibitem{ka1995}
B. Kayser and L. Stodolsky, Phys. Lett. {\bf B 359,} 343-
350 (1995).

\bibitem{an1995}
J. Anandan and H. R. Brown, Foundations of Physics, 25, 349 (1995).

\bibitem{ah1993}
Y. Aharonov, J. Anandan, and L. Vaidman, Physical Review A,
{\bf 47,} no. 6, 4616 (1993); Y. Aharonov
and L. Vaidman,  Physics Letters A, {\bf 178,} 38 (1993); J. Anandan,
Foundations of Physics
Letters, {\bf 6,} No. 6, 503 (1993).

\bibitem{hu1898}
A. Horwitz, Nachr. Gesell. Wiss. G\"ottingen, Math.-Phys. Kl., 309
(1898).

\bibitem{al1947}
A. A. Albert, Ann. Math. {\bf 48,} 495 (1947).

\bibitem{ah1980}
Y. Aharonov and M. Vardi, Phys. Rev. D {\bf 21,} 2235 (1980); J.
Anandan and Y. Aharonov, Phys. Rev. D, {\bf 38,} No. 6, 1863-1870
(1988).

\bibitem{ad1995}
Stephen L. Adler, {\it Quaternionic Quantum Mechanics and Quantum
Fields} (Oxford Univ. Press, 1995), p. 6-7, 109-111.

\bibitem{fe1963}
R.P. Feynman and F.L. Vernon, Ann. Phys. {\bf 24,} 118 (1963); A. Stern, Y.
Aharonov and Y. Imry, Phys. Rev. A {\bf 41,} 3436 (1990);
J. Anandan and Y. Aharonov, Phys. Rev. Lett. {\bf 65,} 1697-1700 (1990).

\bibitem{effective}
David Gross in {\it Recent Developments in Quantum Field Theory},
eds. J. Ambjorn, B.J. Durnhuus \& J.L. Peterson (Elsevier, 1985),
p. 151; Howard Georgi, Annu. Rev. Nucl. Part. Sci. {\bf 43,} 209
(1993); S. Weinberg, Phys. Rev. D {\bf 56,} 2303 (1997).

\bibitem{ah1988}
Y. Aharonov, D. Albert and L. Vaidman, Phys. Rev.
Lett. {\bf 60,} 1351 (1988).

\end{thebibliography}
\end{document}